\documentclass[conference]{IEEEtran}
\IEEEoverridecommandlockouts
\usepackage{cite}
\usepackage{amsmath,amssymb,amsfonts}
\usepackage{algorithmic}
\usepackage{graphicx}
\usepackage{textcomp}
\usepackage{xcolor}
\usepackage{setspace}
\def\BibTeX{{\rm B\kern-.05em{\sc i\kern-.025em b}\kern-.08em
		T\kern-.1667em\lower.7ex\hbox{E}\kern-.125emX}}
\usepackage{subfigure}

\newtheorem{my_lemma}{Lemma}

\title{Performance Analysis of Dual-Hop THz Wireless Transmission for Backhaul Applications}

\author{
	\IEEEauthorblockN{Vinay U. Pai, Pranay Bhardwaj, and S. M. Zafaruddin}\\
	\IEEEauthorblockA{ Department of Electrical and Electronics Engineering, 
		BITS Pilani, Pilani Campus, Pilani-333031, Rajasthan, India\\ Email: \{f20170131, p20200026, syed.zafaruddin\}@pilani.bits-pilani.ac.in}
	
	\thanks{This work was supported in part by the Start-up Research Grant, Department of Science Technology (DST), Science and Engineering Research Board (SERB), India under Start-up Research Grant SRG/2019/002345.}
}

\begin{document}
	\maketitle 
	
\begin{abstract}
THz transmissions suffer from pointing errors due to antenna misalignment and incur higher path loss from the molecular absorption in addition to  the channel fading. In this paper, we employ an amplify-and-forward (AF) dual-hop relaying to mitigate the effect of pointing errors and extend the range of  THz wireless system for  backhaul connectivity. We provide statistical analysis on the performance of the considered system by deriving analytical expressions for the outage probability, average bit-error-rate (BER), average signal-to-noise ratio (SNR), and a lower bound on the ergodic capacity over independent and identical (i.i.d) $\alpha$-$\mu$ fading combined with the statistical effect of pointing errors. Using computer simulations, we validate the derived analysis of the relay-assisted system. We also demonstrate the effect of the system parameters on outage probability and average BER with the help of diversity order. We show that data rates up to several \mbox{Gbps} can be achieved using THz transmissions, which is desirable for next-generation wireless systems, especially for backhaul applications.  
\end{abstract}		

\section{Introduction}
There is currently a lot of research work in both academia, and the industry alike in 5G communication technology that uses millimeter-wave (mmWave) frequency bands up to $60$ GHz \cite{mmW_5g}.  Apart from conventional cellular communication, diverse applications in the field of wireless cognition, autonomous driving, positioning, augmented reality, etc., will need higher performance than what mmWave bands can offer \cite{rappaport}. In cellular networks and Internet of Things (IoT) applications, the increase in the number of wireless devices has put a huge strain on existing communication technologies. There is thus a need to explore high-frequency bands, such as those in the Terahertz (0.3-3 THz) and sub-Terahertz (60-300 GHz) range. These bands offer more flexibility and are also economically more viable than the conventional backhaul infrastructures in difficult terrains and hence will be the frequency bands of choice for future backhaul networks \cite{thz_opt}. 

However, the THz band has its challenges and limitations. While the high-frequency bandwidths can offer much higher rates and lower latency than even 5G systems, signals in these bands suffer from severe path loss and atmospheric attenuation \cite{Kim2015,Kokkoniemi_2018, Wu2020,Sarieddeen2019}. The path loss is higher because of molecular absorption at such small wavelengths. The authors in \cite{Kim2015} have developed an experimental characterization of the THz channel whereas in \cite{Kokkoniemi_2018}, the authors have examined the effect of scattering and absorption losses in the THz band. There is also considerable loss due to antenna misalignment and radio frequency (RF) hardware issues \cite{Boluda_2017,KOKKONIEMI2020}. Hence the physical range of these systems is minimal. Cooperative communication systems are a practical solution to mitigate the impact of path loss and channel conditions \cite{Nosratinia2004}. This motivates us to analyze and explore the performance of such relayed THz systems for the dual-hop scenario.

{\em Related Works:} The dual-hop relaying for THz and heterogeneous networks has also been discussed in the literature \cite{precoding_2020,nano_thz,invivo_thz,outage_thz,mixed_thz_rf,Pranay_2021_TVT}. In \cite{precoding_2020}, the authors derived a closed-form solution for hybrid precoding in two-way relayed THz multiple input multiple output (MIMO) wireless systems. The sum capacity and energy efficiency of the proposed system were better than existing solutions for the THz MIMO relay system. The authors in \cite{nano_thz} investigate the performance of a cooperative relay transmission system for nano-sensor networks in the THz band using both amplify-and-forward (AF) and decode-and-forward (DF) relaying assuming a line of sight (LOS) channel in the THz band. Similarly, in \cite{invivo_thz}, the authors focused on the application of THz communication in in-vivo nano-devices. In \cite{outage_thz}, the authors analyze the outage performance of a dual-hop THz wireless system considering the effect of antenna misalignment error. \cite{mixed_thz_rf} considers a mixed dual-hop case, with an RF and THz link with misalignment error. The small-scale multi-path fading is considered to be a generalized $\alpha$-$\mu$ distribution, which includes Weibull, negative exponential, Nakagami-m, and Rayleigh  fading distributions as special cases. In \cite{Pranay_2021_TVT}, the authors analyzed the performance of a THz-RF wireless link over an $\alpha$-$\mu$ fading channel by deriving closed-form expressions for various performance metrics such as outage probability, average BER, and ergodic capacity. It is desirable to analyze a dual-hop cooperative communication system operating purely with THz links which can be useful for upcoming generations of wireless systems. 

In this paper, we present the performance analysis of a dual-hop THz-THz wireless system by considering generalized $\alpha$-$\mu$ fading combined with misalignment errors using AF protocol. Using the probability density function (PDF) and cumulative distribution function (CDF) for the dual-hop AF relay system, we derive exact closed-form expressions of outage probability, average BER, average SNR, and a lower bound on the ergodic capacity of the relayed system. We also develop diversity order of the system to provide better insights on the system performance at high SNR. We validate  derived analysis using Monte-Carlo simulations and demonstrate the performance of the THz-THz system for backhaul application.

{\em Notations:} Some notations that are used in the paper are as follows. $\Gamma(.)$, $\Gamma(.,.)$, and $\psi(.)$ denote the Gamma, upper incomplete Gamma and digamma function, respectively. $\mathcal{B}_{z} (.,.)$ denotes the Beta function and $ {}_2F_1 (.,.;.;.)$ denotes the Gaussian Hypergeometric function. Finally, $G_{p,q}^{m,n}  \left( x \bigg| \begin{array}{c} a_1, \ldots a_n\\b_1, \ldots b_q\end{array}\right)$ denotes the Meijer's G-function.

\section{System Model}\label{sec:system_model}
We consider a dual-hop relay system in which the information is transmitted from source (S) to destination (D) via a relay (R) node with AF protocol. The considered system may be well suited for backhaul transmissions in small-cell networks, cell-free wireless networks, etc., to transmit the information to the central processing unit (CPU). The THz band facilitates the communication on both the hops, i.e., between source to relay and relay to destination. We assume that direct transmission from the source to the relay is not possible due to obstacles between the source and destination. We consider generalized i.i.d $\alpha$-$\mu$ short-term fading along with zero boresight pointing errors. The received signal at the relay or destination can be represented as 
\begin{equation}
	y_{i} = h_{l_i} h_{pf}{x}_i + w_{i},
\end{equation}
where $i=\{1,2\}$ denote the first hop (S-R) and second hop (R-D), respectively. Here, $ {x}_i $ and $w_i$  are the transmitted signal and additive white Gaussian noise at the $i$th link, and $h_{l_i}$ is the  path-gain of the  THz channel: 	
	\begin{equation}
	h_{l_i} = \frac{c\sqrt{G_{t_i}G_{r_i}}}{4\pi f d_i} \exp\left(-\frac{1}{2}k(f,T,\psi,p)d_i\right)
	\end{equation}
where $k$ is the absorption coefficient \cite{Boulogeorgos_Analytical}. The random variable $h_{pf}$ combines the effect of fading and antenna misalignment having PDF \cite{mixed_thz_rf}:	
	\begin{equation}\label{eq:pdf_hpf}
	f_{|h_{pf}|}(x) = A  x^{\phi-1} \Gamma \left(B, Cx^\alpha \right)
	\end{equation}	
where $\alpha$, $\mu$ are the fading parameters and  $ A = \phi S_0^{-\phi} \frac{\mu^{\frac{\phi}{\alpha}}}{\Omega^\alpha \Gamma (\mu)} $, $ B = \frac{\alpha \mu - \phi}{\alpha} $, and $C =  \frac{\mu}{\Omega^\alpha} S_0^{-\alpha}$. Here, $\Omega$ is the $\alpha$ root mean value of the fading channel envelope and $\phi$, $ S_0 $ are the pointing error parameters as presented in \cite{Farid2007}.	

Using \eqref{eq:pdf_hpf}, the PDF of instantaneous SNR of the THz link can be represented as,
\begin{equation}\label{eq:pdf_thz}
	f_{\gamma_i}(\gamma) = \frac{A}{2\sqrt{\gamma \gamma_0}} \bigg (\sqrt{\frac{\gamma}{\gamma_0}}\bigg)^{\phi-1} \Gamma \bigg(B, C 
	\bigg(\sqrt{\frac{\gamma}{\gamma_0}}\bigg)^\alpha \bigg)
\end{equation}

and the CDF is given as \cite{Pranay_2021_TVT},
\begin{eqnarray}	\label{eq:cdf_thz}
	&F_{\gamma_i}(\gamma)=  \frac{A  C^{-\frac{\phi}{\alpha}}}{\phi} \bigg[ \gamma\bigg(\mu,C\Big(\sqrt{{\gamma}/{\gamma_{0}}}\Big)^{\alpha}\bigg)  \nonumber \\  & + C^{\frac{\phi}{\alpha}}\Big(\sqrt{{\gamma}/{\gamma_{0}}}\Big)^{\phi}\times \Gamma\Big(B,C\Big(\sqrt{{\gamma}/{\gamma_{0}}}\Big)^{\alpha}\Big) \bigg]
\end{eqnarray}

The instantaneous SNR of the THz link is denoted by $\gamma = \gamma_0 |h_{pf}|^2$  where $\gamma_0= {P_i |h_{l_i}|^2}/{\sigma_{w_i}^2}$ is the SNR term without channel fading for the THz link with transmit power $ P_i $. The end-to-end SNR for a channel state information (CSI) assisted AF relay is given by \cite{Hasna_2004_AF}
 \begin{eqnarray}
 \gamma= \frac{\gamma_1\gamma_2}{\gamma_1+\gamma_2+1}
 \end{eqnarray}

\section{Performance Analysis}
In this section, we will derive the analytical expressions of outage probability, average BER, ergodic capacity, and average SNR for the dual-hop relaying system. Since the exact analysis of the  CSI-assisted AF relay demands the use of more complex mathematical functions like Fox's H and rigorous computations, we use an upper bound for the analysis of the considered dual-hop system for which the end to end SNR is given by \cite{papoulis_2002}:
\begin{eqnarray}
	\gamma \leq \min\{\gamma_1,\gamma_2\}
\end{eqnarray}

Thus,  distribution functions are given by
\begin{eqnarray} \label{eq:cdf_relay}
	F_{\gamma}(\gamma) =  F_{\gamma_1}(\gamma)+F_{\gamma_2}(\gamma)-F_{\gamma_1}(\gamma)F_{\gamma_2}(\gamma)
\end{eqnarray}
\begin{eqnarray}\label{eq:pdf_relay}
	f_{\gamma}(\gamma) = f_{\gamma_1}(\gamma)+f_{\gamma_2}(\gamma)-f_{\gamma_1}(\gamma)F_{\gamma_2}(\gamma)-F_{\gamma_1}(\gamma)f_{\gamma_2}(\gamma)
\end{eqnarray}
where $f_{\gamma_1}(\gamma)$, $f_{\gamma_2}(\gamma)$ denote the PDFs of the first and second hops, respectively and $F_{\gamma_1}(\gamma)$ and $ F_{\gamma_2}(\gamma) $ denote the CDFs of the first and second hops, respectively.

\subsection{Outage Probability}\label{sec:perf_anal}
Outage probability is defined as the probability of instantaneous SNR value being less than some threshold value $\gamma_{th}$ such that the system goes into outage i.e., $ P_{\rm out} = P(\gamma <\gamma_{th}) $. The outage probability can be evaluated by substituting $\gamma_{th}$ into \eqref{eq:cdf_relay}, where $ F(\gamma) $ is given in \eqref{eq:cdf_thz}.

The diversity order of the relayed system can be obtained using the asymptotic analysis in high SNR regime \cite{Pranay_2021_TVT}
\begin{eqnarray}\label{eq:diversity order}
	M = \min \bigg\{\frac{\alpha\mu}{2}, \frac{\phi}{2} \bigg\}
\end{eqnarray}

\begin{figure*}
	\small
	\begin{eqnarray} \label{eq:ber}
	&\bar{P}_e \approx \frac{\sqrt{2} \alpha^{(\frac{\alpha(B-1)+2p+\phi-3}{2})} \big(\frac{2q+1}{2}\big)^{-(\frac{\alpha(B-1)+2p+\phi-1}{2})} (C\gamma_0^{\frac{-\alpha}{2}})^{(B-1)}}{(2\pi)^\frac{\alpha}{2}} G_{\alpha,2}^{2,\alpha} \Bigg(\frac{4(C\gamma_0^{\frac{-\alpha}{2}}) \alpha^\alpha 2^{-\alpha}}{(2q+1)^\alpha} \Bigg| \begin{matrix} \Delta(\alpha, \frac{\alpha(B-1)+2p+\phi-1}{2}) \\ \Delta(2,0)
	\end{matrix} \Bigg) \Bigg[\Bigg(\frac{A C^{-\frac{\phi}{2}} q^p} {2\sqrt{2\pi}\phi \Gamma(p)}\Bigg) \nonumber \\ &\times  \left(C^{-\frac{\phi}{2}} \gamma_0^{-\frac{\phi}{2}}\right)+ \left(\frac{A} {2\sqrt{2\pi }\phi \gamma_0^{\frac{\phi}{2}}}\right)^2  \frac{q^p}{\Gamma(p)} 2(C^{-\frac{\phi}{2}} \gamma_0^{-\frac{\phi}{2}})   \Bigg]        +\Bigg(\frac{A C^{-\frac{\phi}{2}} q^p} {2\sqrt{2\pi}\phi \Gamma(p)}\Bigg) \Bigg[\Gamma(\mu) \sqrt{2\pi} \nonumber \\ & -  \sum_{k=0}^{\mu-1} \frac{\Gamma(\mu)(C\gamma_0^{\frac{-\alpha}{2}})^k}{k!} \frac{\sqrt{2} \alpha^{(\frac{2p+\alpha k-3}{2})} \big(\frac{2q+1}{2}\big)^{-(\frac{2p+\alpha k-1}{2})}}{(2\pi)^\frac{\alpha}{2}}  G_{\alpha,2}^{2,\alpha} \Bigg(\frac{4(C\gamma_0^{\frac{-\alpha}{2}})^2 \alpha^\alpha 2^{-\alpha}}{(2q+1)^\alpha} \Bigg| \begin{matrix} \Delta(\alpha, \frac{2p+\alpha k-1}{2}) \\ \Delta(2,0)	\end{matrix} \Bigg)         \Bigg]      + \Bigg(\frac{A} {2\sqrt{2\pi }\phi \gamma_0^{\frac{\phi}{2}}}\Bigg)^2  \frac{q^p}{\Gamma(p)}  \nonumber \\ &  \hspace{-1mm}  \Bigg[\left(C^{\hspace{-1mm}-\frac{\phi}{2}} \gamma_0^{\frac{\phi}{2}}\right)^2 (\Gamma(\mu))^2\sqrt{2\pi}  \hspace{-1mm}+\hspace{-1mm} \frac{\sqrt{2} \alpha^{\frac{2\alpha(B-1)+2p+\phi-3}{2}} \big(\frac{2q+1}{2}\big)^{-\frac{2\alpha(B-1)+2p+\phi-1}{2}} (C\gamma_0^{\frac{-\alpha}{2}})^{2B-2}}{(2\pi)^\frac{\alpha}{2}}  G_{\alpha,2}^{2,\alpha} \Bigg(\hspace{-1mm}\frac{16(C\gamma_0^{\frac{-\alpha}{2}})^2 \alpha^\alpha 2^{-\alpha}}{(2q+1)^\alpha} \Bigg| \begin{matrix} \Delta(\alpha, \frac{2\alpha(B-1)+2p+\phi-1}{2}) \\ \Delta(2,0)
	\end{matrix} \Bigg) \nonumber \\ &  +  \left(C^{-\frac{\phi}{2}} \gamma_0^{\frac{\phi}{2}}\right)^2  \sum_{k_1=0}^{\mu-1} \sum_{k_2=0}^{\mu-1} \hspace{-1mm} \frac{\big(\Gamma(\mu)\big)^2 (C\gamma_0^{\frac{-\alpha}{2}})^{k_1+k_2}}{k_1!k_2!}    \frac{\sqrt{2} \alpha^{(\frac{2p+\alpha k_1+ \alpha k_2-3}{2})} \big(\frac{2q+1}{2}\big)^{-(\frac{2p+\alpha k_1+ \alpha k_2-1}{2})}}{(2\pi)^\frac{\alpha}{2}}\nonumber \\ &  G_{\alpha,2}^{2,\alpha} \Bigg(\frac{4(C\gamma_0^{\frac{-\alpha}{2}})^2 \alpha^\alpha 2^{-\alpha}}{(2q+1)^\alpha} \Bigg| \begin{matrix} \Delta(\alpha, \frac{2p+\alpha k_1+ \alpha k_2-1}{2}) \\ \Delta(2,0)
	\end{matrix} \Bigg)  + 2\big(C^{-\frac{\phi}{2}} 	\gamma_0^{\frac{\phi}{2}}\big)^2 \Gamma(\mu) \sum_{k=0}^{\mu-1} \frac{\Gamma(\mu)(C\gamma_0^{\frac{-\alpha}{2}})^k}{k!} \frac{\sqrt{2} \alpha^{(\frac{2p+\alpha k-3}{2})} \big(\frac{2q+1}{2}\big)^{-(\frac{2p+\alpha k-1}{2})}}{(2\pi)^\frac{\alpha}{2}} \nonumber \\ & G_{\alpha,2}^{2,\alpha} \Bigg(\frac{4(C\gamma_0^{\frac{-\alpha}{2}})^2 \alpha^\alpha 2^{-\alpha}}{(2q+1)^\alpha} \Bigg| \begin{matrix} \Delta(\alpha, \frac{2p+\alpha k-1}{2}) \\ \Delta(2,0)	\end{matrix} \Bigg)  - 2\left(C^{-\frac{\phi}{2}}\gamma_0^{\frac{\phi}{2}}\right) \nonumber\\ &\sum_{k=0}^{\mu-1}\frac{\Gamma(\mu)(C\gamma_0^{\frac{-\alpha}{2}})^{B+k-1}}{k!}\frac{\sqrt{2} \alpha^{(\frac{\alpha k+\alpha(B-1)+2p+\phi-3}{2})} \big(\frac{2q+1}{2}\big)^{-(\frac{\alpha k+\alpha(B-1)+2p+\phi-1}{2})} (C\gamma_0^{\frac{-\alpha}{2}})^{2(B-1)}}{(2\pi)^\frac{\alpha}{2}} \nonumber\\  & \times G_{\alpha,2}^{2,\alpha} \Bigg(\frac{16(C\gamma_0^{\frac{-\alpha}{2}})^2 \alpha^\alpha 2^{-\alpha}}{(2q+1)^\alpha} \Bigg| \begin{matrix} \Delta(\alpha, \frac{\alpha k+\alpha(B-1)+2p+\phi-1}{2})\\ \Delta(2,0)\end{matrix}\Bigg)  \Bigg]
	\end{eqnarray}
	\hrule
\end{figure*}

\subsection{Average BER}
The average BER of the presented system is given as \cite{Ansari2011}
\begin{eqnarray} \label{eq:ber_eqn}
	\bar{P_e} = \frac{q^p}{2\Gamma(p)}\int_{0}^{\infty} \gamma^{p-1} {e^{{-q \gamma}}} F_{\gamma} (\gamma)   d\gamma
\end{eqnarray}
where $p$ and $q$ are modulation parameters.

\begin{my_lemma}
		if $\phi$ and $S_0$ are the pointing error parameters, and $\alpha$, $\mu$ are the fading parameters, then average BER of the relay-assisted link is given in \eqref{eq:ber}.
\end{my_lemma}
\begin{IEEEproof}
	See Appendix A for the proof.
\end{IEEEproof}

It is to be noted that the diversity order for average BER is same as that for the outage probability given in \cite{Pranay_2021_TVT}.

\subsection{Ergodic Capacity}\label{sec:capacity}
Using \eqref{eq:pdf_relay} and the identity $\log_2(1+\gamma)\geq \log_2(\gamma)$, we define a lower bound on the ergodic capacity
\begin{align}\label{eq:rate_eqn}
    \overline{C} =&  \int_{0}^{\infty} \log_2 (\gamma) f_{\gamma} (\gamma) d\gamma
\end{align}

\begin{my_lemma}
if $\phi$ and $S_0$ are the pointing error parameters, and $\alpha$, $\mu$ are the fading parameters, then ergodic capacity of the relay-assisted link is given as
\begin{align}\label{eq:rate}
\overline{C} & = \bigg(\frac{4 A C^{\frac{-\phi}{\alpha}}\Gamma(\mu) \{-\alpha + \phi (\ln{\gamma_0}-\ln{C}+\psi(\mu))\}}{\alpha \phi^3 \ln{2}}\bigg)  \nonumber \\
\times &  (\phi - A C^{\frac{-\phi}{\alpha}}\Gamma(\mu)) + \frac{4 A^2 C^{\frac{-2\phi}{\alpha}} }{\alpha^{2} \phi \ln{2}}  \nonumber \\ \times & \bigg \{ G_{2,1:1,2:2,2}^{0,2:2,0:1,2}\left(\begin{array}{c}1-\frac{\phi}{\alpha}-B,1-\frac{\phi}{\alpha} \\ \frac{-\phi}{\alpha}
\end{array}\bigg| \begin{array}{c} 1\\ \mu, 0\\ \end{array}\bigg| \begin{array}{c} 1,1\\ 1,0\\ \end{array}\bigg| 1,\frac{\gamma_0}{C} \right) \nonumber \\ - G&_{2,1:1,2:2,2}^{0,2:2,0:1,2}\left( \begin{array}{c} 1-\frac{2\phi}{\alpha}-B,1-\frac{2\phi}{\alpha}\\ \frac{-2\phi}{\alpha} \end{array}\bigg| \begin{array}{c} 1\\ B, 0\\
\end{array}\bigg| \begin{array}{c} 1,1\\ 1,0\\ \end{array}\bigg| 1,\frac{\gamma_0}{C} \right) \bigg\}
\end{align} 
\end{my_lemma}.
\begin{IEEEproof}
We substitute \eqref{eq:pdf_thz} in \eqref{eq:rate_eqn}, and define $\overline{C} =  \int_{0}^{\infty} \log_2 (\gamma) 2 (1 - F_{\gamma} (\gamma) )f_{\gamma} (\gamma) d\gamma$.  Further, substituting $\big(\sqrt{{\gamma}/{\gamma_0}}\big)^\alpha$= $t$ we get
\begin{flalign}\label{eq:rate_int}
    &\overline{C} = \frac{4A}{\alpha^2} \Bigg[\int_{0}^{\infty}  \log_2 (\gamma_0t) t^{\frac{\phi}{\alpha} - 1} \Gamma \bigg(B, C t\bigg) dt \nonumber \\ -& \frac{A C^{\frac{-\phi}{\alpha}}}{\phi} \Gamma(\mu) \int_{0}^{\infty}\log_2 (\gamma_0t)  t^{\frac{\phi}{\alpha} - 1}  \Gamma \big(B, C t\big) 
     dt \nonumber \\ -& \frac{A C^{\frac{-\phi}{\alpha}}}{\phi} \int_{0}^{\infty} \log_2 (\gamma_0 ~t) t^{\big(\frac{2\phi}{\alpha} - 1\big)} \Big( \Gamma \big(B, C t\big) \Big)^2  dt \nonumber \\ + & \frac{A C^{\frac{-\phi}{\alpha}}}{\phi}  \int_{0}^{\infty} \hspace{-2mm}  \log_2 (\gamma_0 ~t) t^{\frac{\phi}{\alpha} - 1} \Gamma \big(B, C t\big)
    \Gamma \big(\mu, C t\big)  dt \Bigg]
\end{flalign}

For the first and the second integral, we use the integration-by-parts method with $\log_2 (\gamma_0 t)$ being the first and $t^{\frac{\phi}{\alpha} - 1}  \Gamma (B, C t)$ being the second term. For the third and the fourth integral, we once again make the transformation from $\log_2 (\gamma_0 ~t)$ to $\log_2 (1+ \gamma_0 ~t)$, and transform the $\log(.)$ and $\Gamma(.,.)$ functions to their Meijer's G equivalents \cite{meijer_equi}. Further, applying the identity  \cite[07.34.21.0081.01]{meijer}, we get the closed form expression for these integrals. Finally, by substituting the solutions of integrals in \eqref{eq:rate_int}, we get the analytical expression for lower bound on ergodic capacity as given in \eqref{eq:rate}.
\end{IEEEproof}

\begin{figure*}[tp]
	\begin{center}
		\subfigure[Outage probability analysis for different $\phi$ and fading parameters $(\alpha,\mu)$.]{\includegraphics[scale = 0.35]{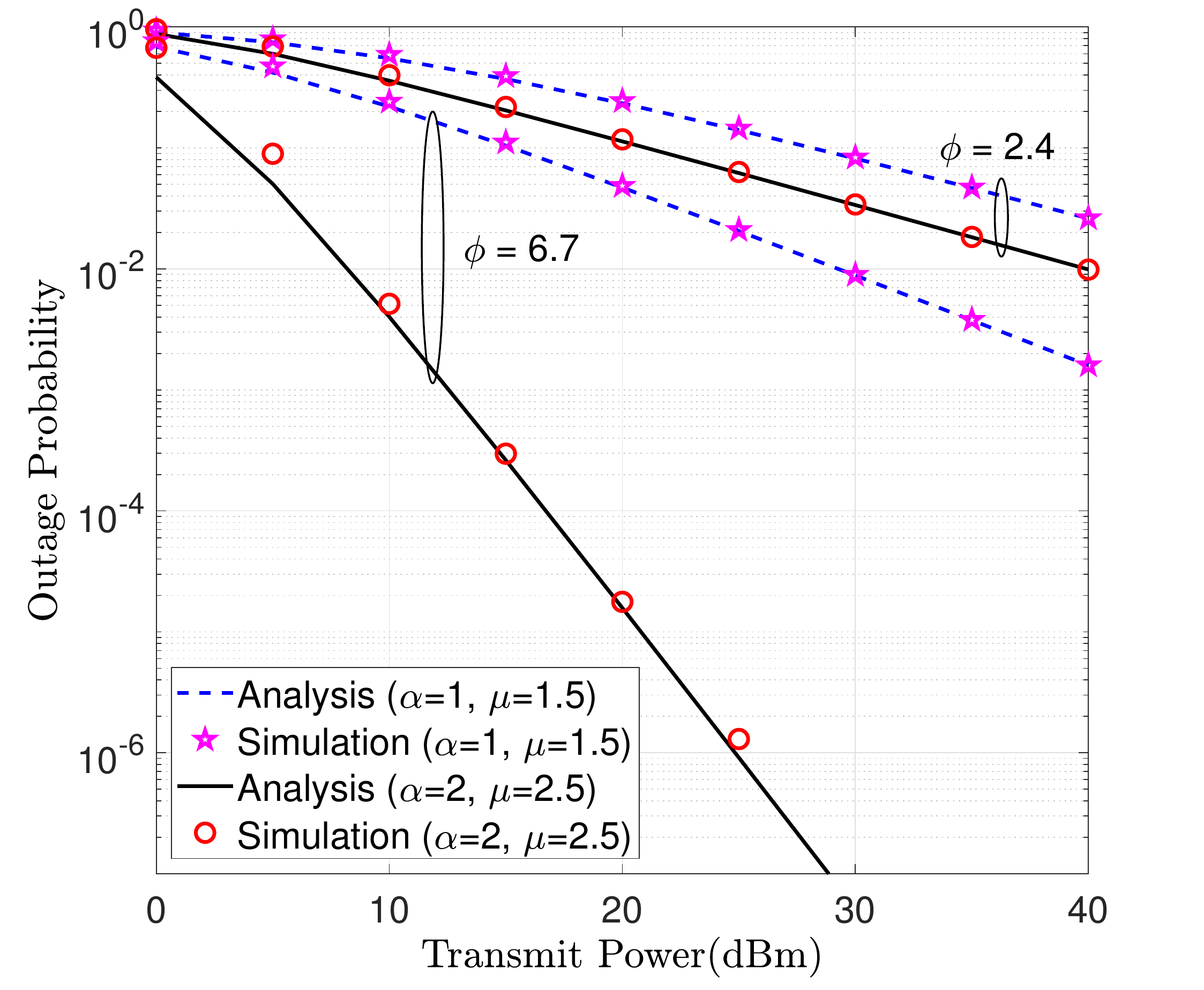}}
		\subfigure[Average BER analysis for different $\phi$ and fading parameters $(\alpha,\mu)$.]{\includegraphics[scale = 0.35]{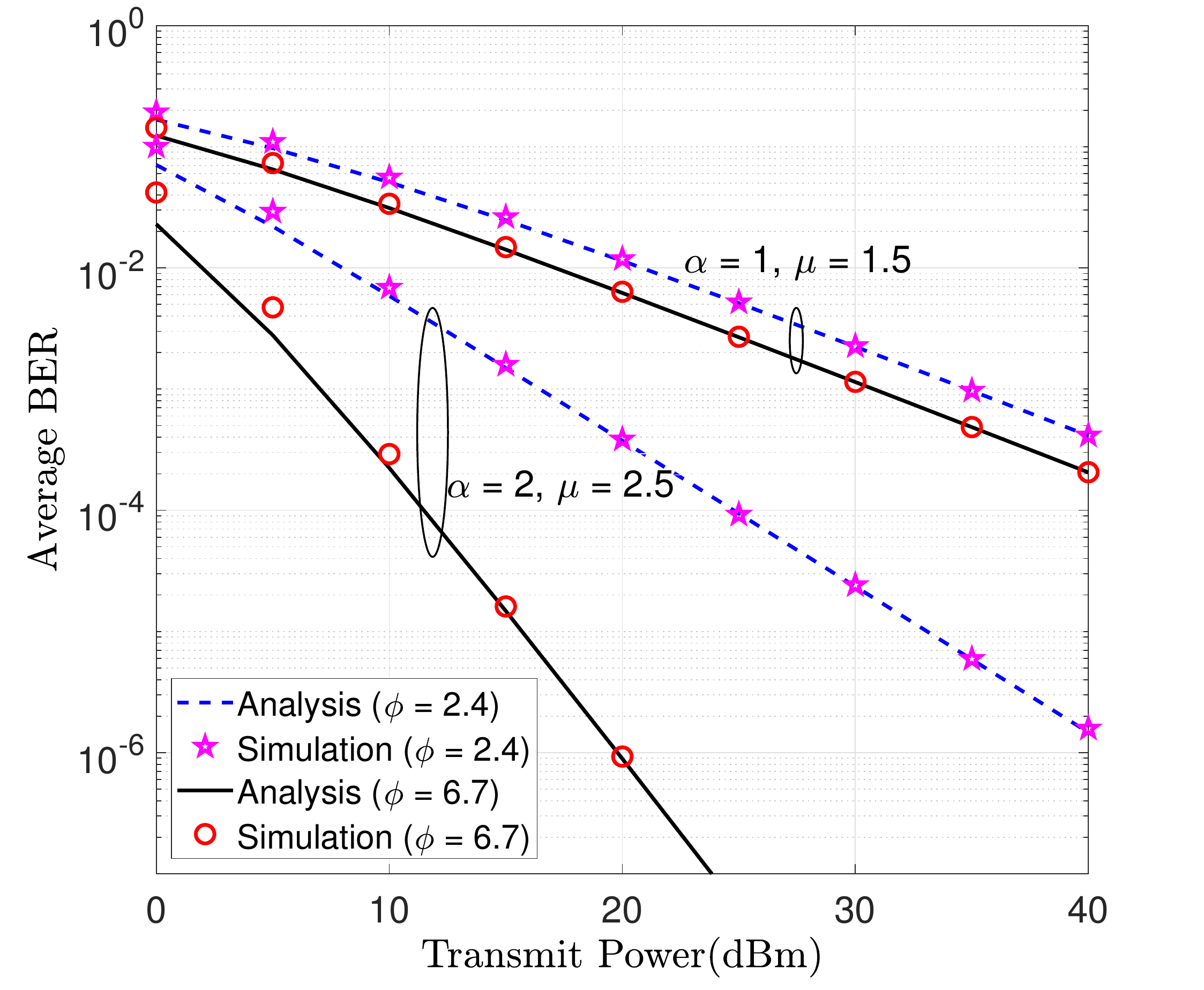}}
		\caption{Outage probability and average BER performance  of relay-assisted THz wireless system.}
		\label{fig:outage_ber}
	\end{center}
\end{figure*}

\subsection{Average SNR}
The average SNR is the expected value of the instantaneous SNR. using \eqref{eq:pdf_relay}, average SNR of the dual-hop system is given as
\begin{flalign}\label{eq:avg_snr_eqn}
\bar{\gamma} =& \int_{0}^{\infty} \gamma f_{\gamma} (\gamma) d\gamma 
\end{flalign}
\begin{my_lemma}
	if $\phi$ and $S_0$ are the pointing error parameters, and $\alpha$, $\mu$ are the fading parameters, then the average SNR of the relay assisted link is given as	
	\begin{eqnarray}\label{eq:avg_snr}
	&\bar{\gamma} = \frac{A \gamma_0 C^{\frac{-2(1+\phi)}{\alpha}}}{\phi (1+\phi)(2+\phi)}
	\nonumber \bigg[ {2(1+\phi)(C^{\frac{\phi}{\alpha}} - A \Gamma(\mu))\Gamma \left(\frac{2}{\alpha} + \mu \right)} \nonumber \\ & +  A \phi \Gamma \big(\frac{2+\alpha \mu + \phi}{\alpha} \big) \Gamma \left(B \right) +  A \Gamma \left(2\left( \frac{1}{\alpha} + \mu \right)  \right)  \nonumber \\ & \times  \Big\{ 2 (-1^{-(\frac{2+\alpha \mu}{\alpha})} ) (1 +\phi) \mathcal{B}_{-1}\left[\frac{2}{\alpha} + \mu, 1 -2\left( \frac{1}{\alpha} + \mu \right) \right] \nonumber \\ & \times  (-1^{(1-B)})\phi \mathcal{B}_{-1}\left[B, 1 -2\big( \frac{1}{\alpha} + \mu \big) \right] \nonumber  \\ & - \left(\frac{\alpha (2+\phi)  {}_2F_1 \big(2( \frac{1}{\alpha} + \mu ),\frac{2+\alpha \mu + \phi}{\alpha}; \frac{2+\alpha + \alpha \mu + \phi}{\alpha};-1  \big)}{2+\alpha \mu + \phi}\right) \bigg\} \bigg] 
	\end{eqnarray} 
\end{my_lemma}

\begin{IEEEproof}	
We substitute \eqref{eq:pdf_relay} in \eqref{eq:avg_snr_eqn}, and define $ \bar{\gamma} = \int_{0}^{\infty} 2 \gamma (1 - F_{\gamma} (\gamma)) f_{\gamma} (\gamma) d\gamma$. Further, substituting $\big(\sqrt{{\gamma}/{\gamma_1^0}}\big)^{\alpha_1} = t$ we get the average SNR of the relay assisted link		
\begin{eqnarray} \label{eq:avg_snr_int}
	&\bar{\gamma} = \frac{2 \gamma_0 A}{\alpha}\bigg[\int_{0}^{\infty}  ~t^{\frac{2+\phi}{\alpha}-1}  \Gamma \big(B, C t\big) dt - \frac{A C^{\frac{-\phi}{\alpha}}}{\phi} \Gamma(\mu)  \nonumber \\ & \int_{0}^{\infty}   t^{\frac{2+\phi}{\alpha}-1}  \Gamma \big(B, C t\big) 
	dt
	 -\frac{A}{\phi}\int_{0}^{\infty}  t^{\frac{2(1+\phi)}{\alpha}-1}   \big( \Gamma \big(B, C t\big) \big)^2 dt \nonumber \\ & + \frac{A C^{\frac{-\phi}{\alpha}}}{\phi} \Gamma(\mu)  \int_{0}^{\infty}   t^{\frac{2}{\alpha}}  t^{\frac{\phi}{\alpha} - 1}  \Gamma \big(B, C t\big)  \Gamma \big(\mu, C t\big) \bigg] dt  
\end{eqnarray}
	
The first and the second integral can be solved using the identity \cite[6.455/1]{integrals}. The third integral is solved by applying the integration-by-parts method with $\Gamma \big(B, C t\big)$ as the first and $t^{\frac{2(1+\phi)}{\alpha}-1} \Gamma \big(B, C t\big)$ as the second term. We follow a similar procedure to solve the fourth integral. Finally, using the limits of the integrals in the following identity, we get the solution for both the integrals.
\begin{equation}
	\int t^{x-1} \Gamma(a,t) dt = \frac{t^x \Gamma(a,t) - \Gamma(a+x,t)}{x}
\end{equation}
	
Further, to simplify the expression, we use the following identity to represent hypergeometric functions,
\begin{equation}
	{B}_{z}(a,b) = \frac{z^a }{a}  {}_2F_1 (a,1-b;a+1;z)
\end{equation}
	
By solving the integrals and substituting them into \eqref{eq:avg_snr_int}, we get the closed form expression for the average SNR as given in \eqref{eq:avg_snr}.
\end{IEEEproof}

\section{Simulation and Numerical Results}\label{sec:sim_results}

\begin{figure*}[tp]
	\begin{center}
		\subfigure[Average SNR for different values of $\phi$, $\alpha=1$ and $\mu=1.5$.]{\includegraphics[scale = 0.35]{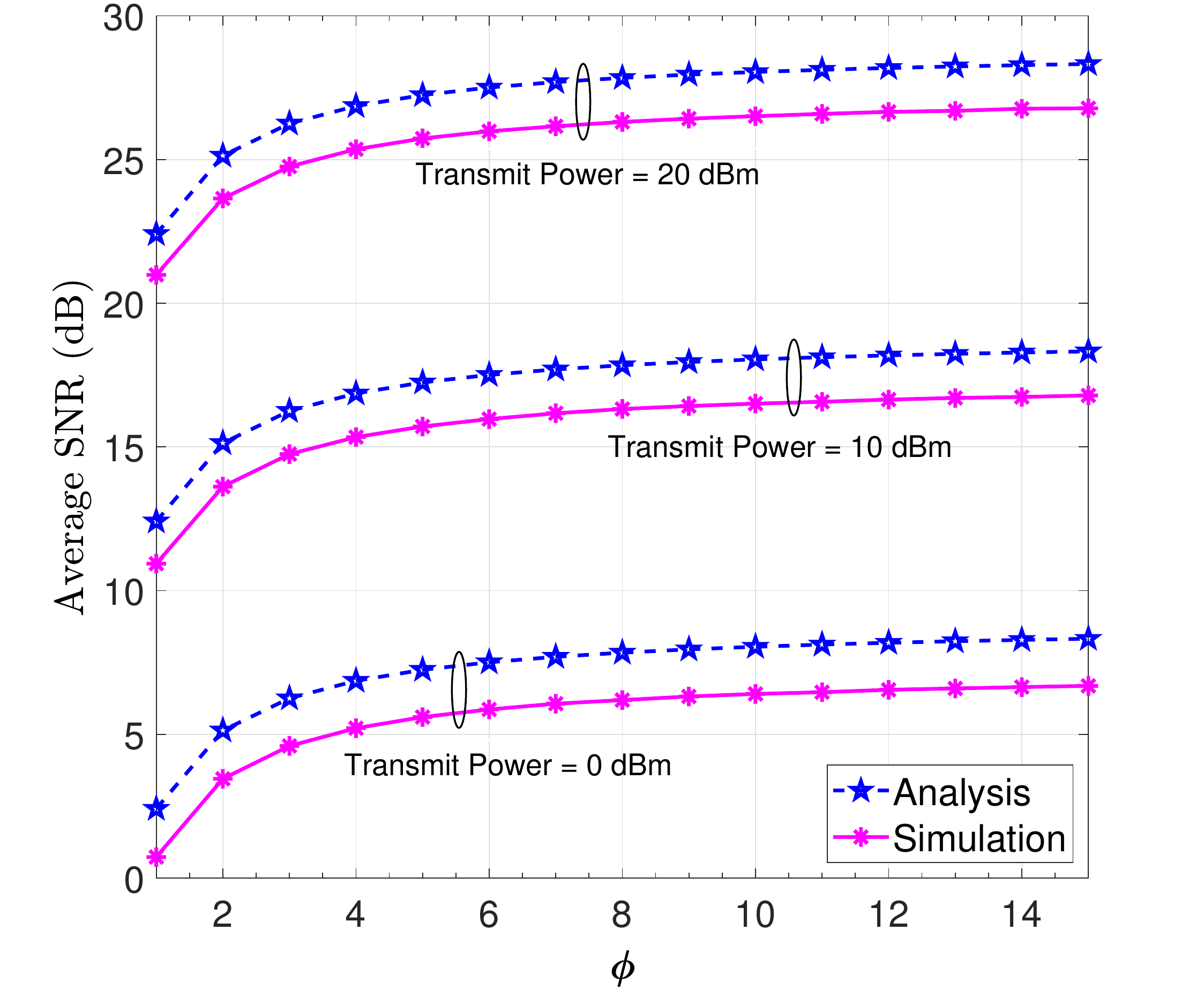}}
		\subfigure[Total capacity for different link distances, $\phi=2.4$, $\alpha=1$ and $\mu=1.5$.]{\includegraphics[scale = 0.35]{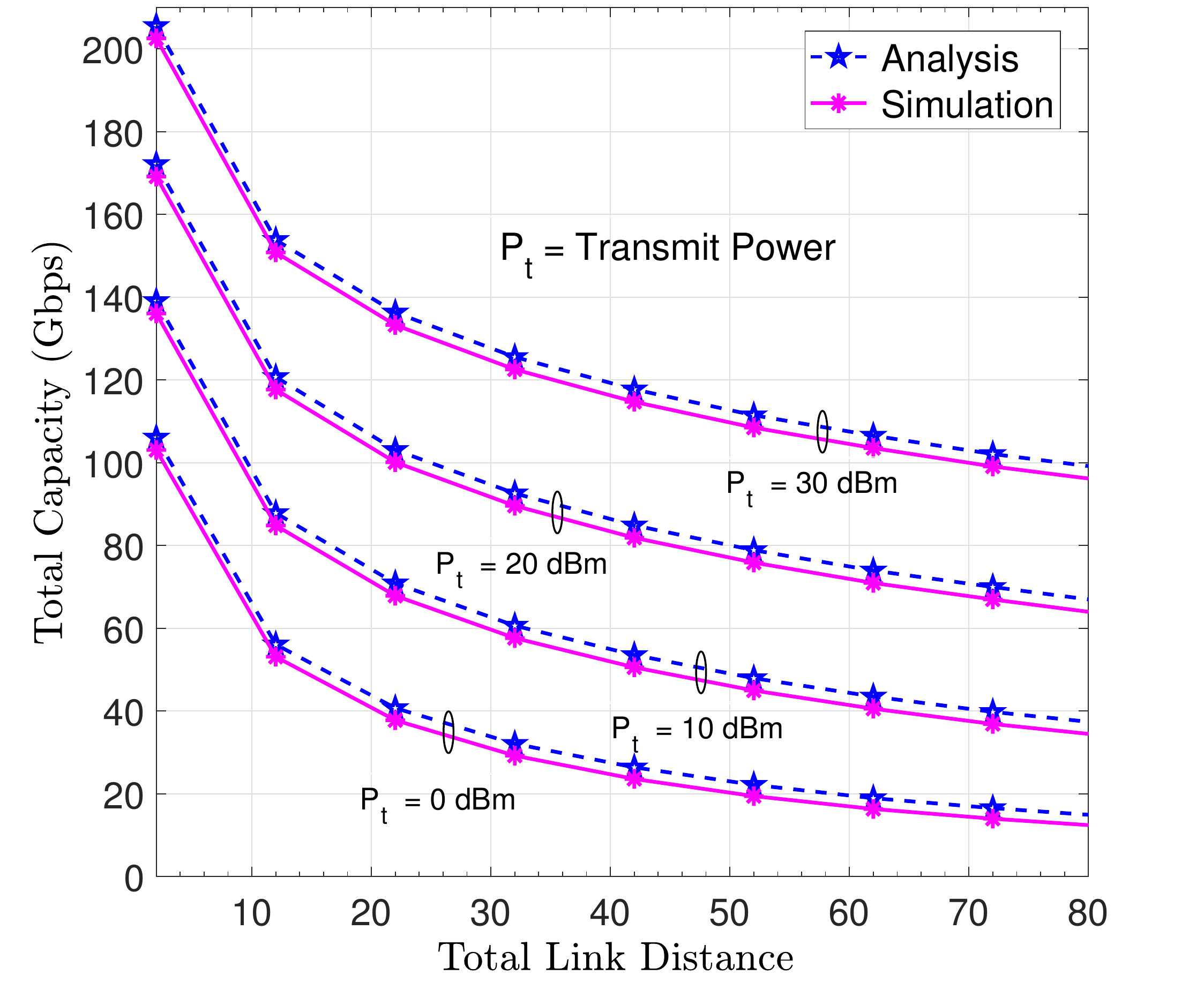}}
		\caption{Average SNR and data rate performance of relay-assisted THz wireless system.}
		\label{fig:snr_rate}
	\end{center}
\end{figure*}

In this section, we validate the derived analytical expressions with the help of Monte-Carlo simulations (averaged over $10^7$ channel realizations) using  MATLAB software. We consider an operating frequency of $ 275 $ \mbox{GHz} with transmit and receive antenna gains as $45$ \mbox{dBi}.  The path-loss is calculated for the THz link using the parameters given in \cite{Boulogeorgos_Analytical}. The value of fading parameters $\alpha$ and $\mu$ are taken in the range of $1-3$ to model the generalized $\alpha$-$\mu$ distribution. Pointing error parameters $\phi$ and $S_0$ are calculated using the values provided in \cite{Farid2007}. We take the noise PSD for THz to be $-174$ \mbox{dBm} and the channel bandwidth as $10$ \mbox{GHz} \cite{Sen_2020_Teranova}.

In Fig. \ref{fig:outage_ber}a, we illustrate the outage probability performance of the relayed system for an SNR threshold of $4$ \mbox{dB}. It is evident from the figure that the outage performance decreases with an increase in parameter $\phi$, for the same $\alpha$-$\mu$ fading parameter. We can observe the impact of the fading and pointing error parameters on the diversity order for the outage performance. The diversity order depends on the pointing error parameter for different values of fading parameters $\alpha$ and $\mu$ when $\phi=6.7$, which can be verified by the change in slope of the plots. In Fig. \ref{fig:outage_ber}b, we present the average BER performance of the system. Similar to the outage probability, the average BER decreases when the parameter $\phi$ increases. When the values of the channel parameters $\alpha$=1 and $\mu$=1.5, the diversity order remains unchanged, thus, the slope of the BER plot does not change. However, when $\alpha$ and $\mu$ are increased to $2$ and $2.5$ respectively, the diversity order  becomes dependent on $\phi$ and a change in the slope of  plots confirms the diversity order analysis. 

In Fig. \ref{fig:snr_rate}a, we demonstrate the average SNR performance of the relaying system. The average SNR is plotted  versus pointing error parameter $\phi$ for different transmitted power, and $\alpha = 1$ and $\mu = 1.5$. As $\phi$ is increased, the average SNR first increases and then becomes constant, which implies that there is not much improvement in the link performance beyond a certain value of $\phi$. We can also see a significant increase in the average SNR as the transmitted power is increased. Finally, in Fig. \ref{fig:snr_rate}b, we analyze the ergodic capacity of the dual-hop system over a bandwidth of $10$ \mbox{GHz}. As expected, the capacity decreases when the link distance increases. Further, an increase in the  transmit power from $0$ dBm to $30$ dBm increases the ergodic capacity by nearly $100$ Gbps. Fig. \ref{fig:snr_rate}b also provides valuable insights on the link distance versus data rate performance. A high data rate of $100$ Gbps is achievable even when the link distance is $40$ \mbox{m} with a transmit power of $30$ dBm. Thus,  different  capacity requirement for backhaul applications  can be achieved by configuring the dual-hop system with appropriate transmit power and link distance.

\section{Conclusion and Future Work}\label{sec:conc}
In this paper, we investigated the performance of a dual-hop THz-THz wireless system. We considered a generalized i.i.d $\alpha$-$\mu$ fading channel model combined with the statistical pointing errors. We  derived exact analytical expressions for the outage probability, average BER, average SNR, and a lower bound on the ergodic capacity of the relay-assisted system.  The derived results were verified with  Monte-Carlo simulations under different channel conditions and system parameters. We also  verified the diversity order of the system to provide insights on the system performance at a high SNR.  The ergodic capacity analysis demonstrates  that data rates up to several \mbox{Gbps} can be achieved with the THz transmissions, which may fulfill the demands of  future generation wireless systems. The work presented in this paper can be extended with an exact analysis on the performance of CSI-assisted AF relay for the  dual-hop THz-THz system with hardware impairments. 

\section*{Appendix A: Proof of Lemma 1}
Using \eqref{eq:cdf_relay} in \eqref{eq:ber_eqn}, we define the average BER as $\bar{P_e} = \bar{P_e}_{I_1}+\bar{P_e}_{I_2}$, where $\bar{P_e}_{I_1} = \frac{2q^p}{\Gamma(p)} \int_{0}^{\infty} \gamma^{p-1} e^{\frac{q\gamma}{2}}  F_\gamma(\gamma) d\gamma$, which can be simplified as 
\begin{eqnarray} \label{eq:I_1_int}
&\bar{P_e}_{I_1} = \left(\frac{A C^{-\frac{\phi}{2}} q^p} {2\sqrt{2\pi}\phi \Gamma(p)}\right) \bigg[\int_{0}^{\infty} \Gamma(\mu) e^{-\gamma \big(\frac{2q+1}{2}\big)} \gamma^{\frac{2p-3}{2}} d\gamma  \nonumber \\ &+ \Big({C^{-\frac{\phi}{2}} \gamma_0^{-\frac{\phi}{2}}}\Big) \int_{0}^{\infty} \Gamma\big(B,C\big(\frac{\gamma}{\gamma_0}\big)^{\frac{\alpha}{2}}\big) e^{-\gamma \big(\frac{2q+1}{2}\big)} \gamma^{\frac{2p-3}{2}} d\gamma \nonumber \\& - \int_{0}^{\infty} \Gamma\big(\mu,C\big(\frac{\gamma}{\gamma_0}\big)^{\frac{\alpha}{2}}\big) e^{-\gamma \big(\frac{2q+1}{2}\big)} \gamma^{\frac{2p-3}{2}} d\gamma   \bigg]
\end{eqnarray}

The first integral is straight forward and can be solved easily. For the second integral we use the approximation for  $\Gamma\big(B,C {\gamma_{0}}^{\frac{-\alpha}{2}} \gamma^{\frac{\alpha_1}{2}} \big) $ using $\Gamma(a,x) \approx e^{-x}x^{a-1}$. Further, we apply the identity of the product of two Meijer's G-function \cite{meijer} to get the solution. For the third integral, we use the series expansion of ${\Gamma\big(\mu, B {\gamma_{0}}^{\frac{-\alpha}{2}} \gamma^{\frac{\alpha}{2}}\big)}$ using $\Gamma(a,bx)$ =$(a-1)! e^{(-bx)}$ $\sum_{k=0}^{a-1}\frac{(bx)^k}{k!}$ and apply the identity \cite{meijer}. We define $\bar{P_e}_{I_2} = \frac{q^p}{\Gamma(p)} \int_{0}^{\infty} \gamma^{p-1} e^{\frac{q\gamma}{2}}  \big(F_\gamma(\gamma)\big)^2 d\gamma$, which can be rewritten as 
\begin{flalign}\label{eq:I_2_int}
&\bar{P_e}_{I_2}\hspace{-1mm} = \hspace{-1mm}\Bigg(\hspace{-1mm}\frac{A} {\sqrt{8\pi }\phi \gamma_0^{\frac{\phi}{2}}}\hspace{-1mm}\Bigg)^2 \hspace{-2mm}\frac{q^p}{\Gamma(p)}  \Bigg[\hspace{-1mm}\Big(\hspace{-1mm}{C^{-\frac{\phi}{2}} \gamma_0^{\frac{\phi}{2}}}\Big)^{\hspace{-1mm}2}  \hspace{-2mm}\int_{0}^{\infty} \hspace{-4mm} {(\Gamma(\mu))}^2 e^{-\gamma \big(\frac{2q+1}{2}\big)} \gamma^{\frac{2p-3}{2}} d\gamma  \nonumber \\& +   \int_{0}^{\infty} \bigg(\Gamma\big(B,C\big(\frac{\gamma}{\gamma_0}\big)^{\frac{\alpha}{2}}\big)\bigg)^2 e^{-\gamma \big(\frac{2q+1}{2}\big)} \gamma^{\frac{2p-3}{2}} d\gamma \nonumber \\ &+ \Big({C^{-\frac{\phi}{2}} \gamma_0^{\frac{\phi}{2}}}\Big)^2   \int_{0}^{\infty} \bigg( \Gamma\big(\mu,C\big(\frac{\gamma}{\gamma_0}\big)^{\frac{\alpha}{2}}\big)\bigg)^2 e^{-\gamma \big(\frac{2q+1}{2}\big)} \gamma^{\frac{2p-3}{2}} d\gamma  \nonumber \\ &+ 2\Big({C^{-\frac{\phi}{2}} \gamma_0^{\frac{\phi}{2}}}\Big)^2 \int_{0}^{\infty} \Gamma(\mu) \Gamma\big(\mu,C\big(\frac{\gamma}{\gamma_0}\big)^{\frac{\alpha}{2}}\big) e^{-\gamma \big(\frac{2q+1}{2}\big)} \gamma^{\frac{2p-3}{2}} d\gamma  \nonumber \\ &+ 2\Big({C^{-\frac{\phi}{2}} \gamma_0^{\frac{\phi}{2}}}\Big)  \int_{0}^{\infty}  \Gamma(\mu) \Gamma\big(B,C\big(\frac{\gamma}{\gamma_0}\big)^{\frac{\alpha}{2}}\big) e^{-\gamma \big(\frac{2q+1}{2}\big)} \gamma^{\frac{2p-3}{2}} d\gamma  \nonumber \\ &-\hspace{-1mm} 2\Big(\hspace{-1mm}{C^{\hspace{-1mm}-\frac{\phi}{2}} \gamma_0^{\frac{\phi}{2}}}\Big)  \hspace{-2mm}\int_{0}^{\infty} \hspace{-4mm} \Gamma\big(\mu,C\big(\frac{\gamma}{\gamma_0}\big)^{\hspace{-1mm}\frac{\alpha}{2}}\big)  \Gamma\big(B,C\big(\frac{\gamma}{\gamma_0}\big)^{\hspace{-1mm}\frac{\alpha}{2}}\big) e^{\hspace{-1mm}-\gamma \big(\hspace{-1mm}\frac{2q+1}{2}\hspace{-1mm}\big)} \hspace{-1mm}\gamma^{\frac{2p-3}{2}} d\gamma  \Bigg]
\end{flalign}
following the similar procedure applied for \eqref{eq:I_1_int}, we solve \eqref{eq:I_2_int}. 

The first integral is straight forward and can be solved easily. For second integral we use approximation for  $\Gamma\big(B,C {\gamma_{0}}^{\frac{-\alpha}{2}} \gamma^{\frac{\alpha_1}{2}} \big) $ using $\Gamma(b,y) \approx e^{-y}y^{b-1}$. Further, we apply the identity \cite[07.34.21.0012.01
]{meijer} to get the solution. For solving the third integral, we use the series expansion for ${\Gamma\big(\mu, B {\gamma_{0}}^{\frac{-\alpha}{2}} \gamma^{\frac{\alpha}{2}}\big)}$ using $\Gamma(p,by)$ =$(p-1)! e^{(-by)}$ $\sum_{k=0}^{p-1}\frac{(by)^k}{k!}$ and apply the identity \cite{meijer}. Likewise, following the similar procedure for subsequent integrals and upon adding the solutions we get \eqref{eq:ber}.

\bibliographystyle{IEEEtran}
\bibliography{THz,ANTS,Thz_references_others}

\end{document}